\documentclass[conference]{IEEEtran}

\IEEEoverridecommandlockouts                              

\usepackage{amsmath,amssymb,amsbsy,amsfonts,amsthm,latexsym,
               amsopn,amstext,amsxtra,euscript,amscd,color,commath,mathtools}
\usepackage{graphicx}
\usepackage{float}
\usepackage{braket}
\usepackage{dcolumn}
\usepackage{tabularx}
\usepackage{longtable}
\usepackage{float}
\usepackage{epsfig}
\usepackage{comment}
\usepackage{array}
\usepackage{pdfpages}

\def\cH{{\mathcal H}}

\def\cW{{\mathcal W}}

\def\F{{\mathbb F}}
\def\C{{\mathbb C}}
\def\Z{{\mathbb Z}}

\def\Q{{\mathbb Q}}
\def\R{{\mathbb R}}
\def\F{{\mathbb F}}

\def\Z{{\mathbb Z}}

\def\e{{\mathbf e}}

\title{\LARGE \bf
Generalized Boolean Functions and Quantum Circuits
on IBM-Q}

\author{\IEEEauthorblockN{Sugata Gangopadhyay}
\IEEEauthorblockA{\textit{Department of Computer Science and Engineering} \\
\textit{Indian Institute of Technology Roorkee}\\
Roorkee, INDIA \\
sugatfma@iitr.ac.in}
\and
\IEEEauthorblockN{Vishvendra Singh Poonia}
\IEEEauthorblockA{\textit{Department of Electronics and Communication Engineering} \\
\textit{Indian Institute of Technology Roorkee}\\
Roorkee, INDIA \\
vspfec@iitr.ac.in}
\and
\IEEEauthorblockN{\hspace{1.75cm} Daattavya Aggarwal}
\IEEEauthorblockA{\textit{\hspace{1.75cm} Department of Physics} \\
\textit{\hspace{1.75cm} Indian Institute of Technology Roorkee}\\
\hspace{1.75cm} Roorkee, INDIA \\
\hspace{1.75cm} daggarwal@ph.iitr.ac.in}
\and
\IEEEauthorblockN{\hspace{1.5cm} Rhea Parekh}
\IEEEauthorblockA{\textit{\hspace{1.5cm} Department of Physics} \\
\textit{\hspace{1.5cm} Indian Institute of Technology Roorkee}\\
\hspace{1.5cm} Roorkee, INDIA \\
\hspace{1.5cm} rheaparekh@ph.iitr.ac.in}
}

\begin{document}
\vspace*{\fill}
\begin{center}
\begin{minipage}{1\textwidth}
\paragraph{\textbf{IEEE Copyright Notice}}
\textsuperscript{\textcopyright} 2019 IEEE. Personal use of this material is permitted. However, permission to reprint/republish this material for advertising or emotional purposes or for creating new collective works for resale or redistribution to servers or lists or to reuse any copyrighted component of this work in other works,must be obtained from the IEEE. Contact Manager, Copyrights and Permissions/ IEEE Service Center/ 445 Hoes Lane/ P.O.Box 1331/ Piscataway, NJ 08855-1331, USA. Accepted to Publish in: 10th International Conference on Computing. Communication and Networking Technologies (ICCCNT 2019, scheduled for July 6-8 (2019) in Kanpur, UP, India 
\end{minipage}
\end{center}
\vfill
\maketitle
\thispagestyle{empty}
\pagestyle{empty}
\begin{abstract}
We explicitly derive a connection between quantum circuits utilising IBM's quantum gate set and  multivariate quadratic polynomials over integers modulo $8$. We demonstrate that the action of a quantum circuit over input qubits can be written as generalized Walsh-Hadamard transform.  
Here, we derive the polynomials corresponding to
implementations of the Swap gate and Toffoli gate using IBM-Q gate set. 
\end{abstract}
\vspace{2mm}

\section{INTRODUCTION}

IBM quantum experience provides web access to the quantum computer of IBM~\cite{IBMQ}. 
In this paper we adapt the theoretical framework proposed by Montanaro \cite{Montanaro2017} for basic gates used by the IBM quantum computer 
and demonstrate the relation between quantum circuits prepared by using those 
gates and multivariate polynomials over integers modulo $8$. As a particular 
example we derive a polynomial corresponding to a standard implementation of swap gate and 
Toffoli gate on IBM quantum computer.
Recently, Montanaro~\cite{Montanaro2017} showed that a quantum circuit made of Hadamard, controlled-Z, and controlled-controlled-Z gates corresponds to a degree-3 boolean polynomial and the amplitude of the quantum circuit is equal to the difference between number of zeroes and ones of that polynomial. Montanaro's work was inspired from Dawson et. al.~\cite{Dawson2005} who used a different universal gate set. On the similar lines, Rudolph~\cite{Rudolph2009} presented a framework to encode quantum circuit amplitude in terms of matrix permanent. Datta et. al.~\cite{Datta2005} in their work studied trace of the unitary operator in terms of zeroes of a degree-3 polynomial.
Such abstractions of quantum circuits facilitate the analysis of computational tasks on these circuits. 
We believe that further investigations in this direction may lead to compact representations of quantum circuits as well as development of quantum circuit optimization strategies including error minimization in executing a given computational task. 

\section{Preliminaries}

\subsection{Generalized Boolean functions}

Let $\C$, $\R$, $\Q$ be the fields of complex numbers, real numbers, rational numbers,
respectively, and  $\Z$ be the ring of integers. 
A complex number $z \in \C$ can be written as $z = \mathfrak{R}(z)  + \imath \mathfrak{I}(z)$ 
where $\mathfrak{R}(z), \mathfrak{I}(z) \in \R$ are the real and the imaginary parts of $z$, 
respectively.  
The conjugate of $z$ is $\bar z = \mathfrak{R}(z) - \imath \mathfrak{I}(z)$, and the 
modulus of $z$ is $\abs{z} = z \bar z = \mathfrak{R}(z)^2 + \mathfrak{I}(z)^2$. 
Let $\Z_q$ be 
the ring of integers modulo $q$ where $q \in \Z^+$, the set of positive integers, 
and $[n] = \{i : i \in \Z^+, 1\leq i \leq n\}$. 
Let $\F_2$ be the prime field of characteristic $2$, and 
$\F_2^n = \{x=(x_1, \ldots, x_n) : x_i \in \F_2, \mbox{ for all } i \in [n]\}$ 
is the $n$-dimensional vector space over $\F_2$. Hamming weight (or, weight) of a vector 
$x \in \F_2^n$ is ${\rm wt}(x) = \sum_{i\in[n]} x_i$ where the sum is over 
$\Z$. A function $f : \F_2^n \rightarrow \F_2$ is said to be a Boolean function 
in $n$ variables. The algebraic normal form (ANF) of $f$ is 
\begin{equation}
\label{intro-eq1}
f(x_1, \ldots, x_n) = \sum_{a=(a_1, \ldots, a_n) \in \F_2^n} \mu_a x_1^{a_1}\ldots x_n^{a_n}
\end{equation}
where $\mu_a \in \F_2$, for all $a \in \F_2^n$. The set of all $n$-variable Boolean 
functions is denoted by $\mathfrak{B}_n$. The (algebraic) degree of $f \in \mathfrak{B}_n$ 
is  $\deg(f) = \max \{ {\rm wt}(a) : \mu_a \neq 0\}$. 
Boolean functions of degree at most $1$ are said to be affine Boolean functions. 
An inner product of two vectors $x = (x_1, \ldots, x_n), y = (y_1, \ldots, y_n) \in \F_2^n$
is defined as $x \cdot y =  \sum_{i\in [n]} x_iy_i$. For every 
$a \in \F_2^n$, $\epsilon \in \F_2$ we define $\varphi_{a, \epsilon} \in \mathfrak{B}_n$ by 
\begin{equation}
\label{intro-eq2}
\varphi_{a,\epsilon}(x) = a \cdot x + \epsilon, \mbox{ for all } x \in \F_2^n.
\end{equation}
It is known that the set of all $n$-variable affine Boolean functions is
$\mathfrak{A}_n = \{\varphi_{a, \epsilon} : a \in \F_2^n, \epsilon \in \F_2 \}$. 
The functions of the form $\varphi_{a, 0}$ are said to be linear functions, and 
will be denoted by $\varphi_a$. 
The Walsh-Hadamard transform of $f \in \mathfrak{B}_n$ at $u \in \F_2^n$ is 
\begin{equation}
\label{intro-eq3}
\cW_f(u) = \sum_{x \in \F_2^n} (-1)^{f(x) + u \cdot x}.
\end{equation}
The multiset $[\cW_f(u) : u \in \F_2^n ]$ is said to be the Walsh-Hadamard 
spectrum of $f$. Each element of this spectrum $\cW_f(u)$ is referred to 
as a Walsh-Hadamard coefficient of $f$. The Walsh-Hadamard coefficients
satisfy Parseval's identity
\begin{equation}
\label{intro-eq4}
\sum_{u\in \F_2^n} \cW_f(u)^2  = 2^{2n}. 
\end{equation} 

There are several possible generalizations of Boolean functions. We consider
one specific generalization where instead of considering functions from 
$\F_2^n$ to $\F_2$ we consider functions from $\F_2^n$ to $\Z_q$ where
$q \in \Z^+$. Let the set of all functions from $\F_2^n$ to $\Z_q$ be 
denoted by $\mathfrak{B}_n^q$. The generalized Walsh-Hadamard transform of 
$f$ at $u\in \F_2^n$ is 
\begin{equation}
\label{intro-eq5}
\cH_f^{(q)}(u) = \sum_{x \in \F_2^n} \zeta_q^{f(x)}(-1)^{u \cdot x}
\end{equation}
where $\zeta_q = \exp(\frac{2\pi \imath}{q})$ is a primitive $q$th root of unity. 
As before, these are also known as generalized Walsh-Hadamard coefficients. 
The generalized Walsh-Hadamard spectra is $[\cH_f^{(q)}(u) : u \in \F_2^n ]$,
and the Parseval's identity takes the form 
\begin{equation}
\label{intro-eq6}
\sum_{u\in \F_2^n} \abs{\cH_f^{(q)}(u)}^2  = 2^{2n}. 
\end{equation} 
The set of all generalized Boolean functions in $n$ variables from 
$\F_2^n$ to $\Z_q$ is denoted by is denoted by $\mathfrak{GB}_n^q$. For initial 
studies on this generalization we refer to the articles of Sol\'e and Tokareva \cite{SOLE}
and St\u anic\u a, Martinsen, Gangopadhyay and Singh \cite{GAN}. Defining 
algebraic normal form for generalized Boolean functions is not as straight forward 
as in the case of Boolean functions. 

\subsection{IBM's quantum computer}
\label{ibm}
 
The quantum computer made available by IBM as a part of 
IBM Quantum Experience consists of the six single-qubit gates and one 
two-qubit gate, as follows.
\begin{center}
$\begin{array}{lll}
X=
  \begin{bmatrix}
     0&1\\
     1&0\\
  \end{bmatrix}, &
Y=
  \begin{bmatrix}
      0&-\imath\\
      \imath &0\\
  \end{bmatrix},
&
Z=
  \begin{bmatrix}
     1&0\\
     0&-1\\
  \end{bmatrix},

\end{array}$
\end{center}

\begin{center}

$\begin{array}{ll}
S=
  \begin{bmatrix}
     1&0\\
     0&\exp(\frac{\imath \pi}{2})\\
  \end{bmatrix},
&
T=
  \begin{bmatrix}
     1&0\\
     0& \exp(\frac{\imath \pi}{4})\\
  \end{bmatrix},
\end{array}$
\end{center}

\begin{center}
$\begin{array}{ll}
H=
\frac{1}{\sqrt{2}}
  \begin{bmatrix}
     1&1\\
     1&-1\\
  \end{bmatrix},
  &
CNOT=
    \begin{bmatrix}
     1&0&0&0\\
     0&1&0&0\\
     0&0&0&1\\
     0&0&1&0\\
   \end{bmatrix}.
\end{array}$
\end{center}

This gate set is universal \cite{Boykin2000}, i.e., any quantum circuit (algorithm) can be realized by 
using these gates exclusively. 
Montanaro \cite{Montanaro2017} proposed a method to connect Boolean functions 
to quantum circuits consisting of $Z$, $CZ$, and $CCZ$ gates. In the next section 
we employ similar consideration to construct Boolean functions corresponding to the 
IBM quantum experience (IBM-Q) gate-set. We demonstrate that for any circuit 
on IBM-Q there exist a generalized Boolean function from $\F_2^n$ to $\Z_8$  \cite{Dawson2005}
whose generalized Walsh--Hadamard spectrum tells the probability amplitudes 
of the measurement with respect to the standard basis.

\section{Deriving generalized Boolean functions from quantum circuits}

\begin{figure}[h!]
\begin{center}
\includegraphics[width= 0.4\linewidth]{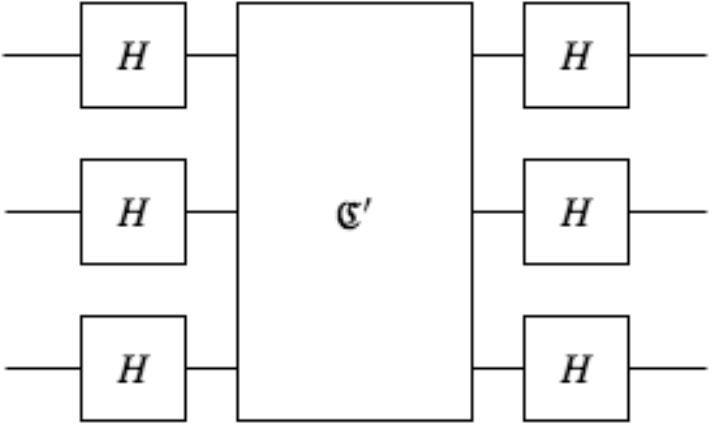}
\end{center}
\caption{The quantum circuit $\mathfrak C$.}
\label{fig1}
\end{figure}

Suppose $\mathfrak{C}$ is a quantum circuit represented in Fig.~\ref{fig1}. 
Each horizontal line corresponds to a qubit. 
It is known that we can add pairs of Hadamard gates to the beginning or end of each line without changing the unitary operator corresponding to the circuit~\cite{Montanaro2017}. Thus we will consider the inner circuit $\mathfrak{C}'$. We also assume that there is at least one gate acting on each qubit. Consider an $n$-qubit quantum state represented as a $\C$-linear combination of the vectors of the standard basis. 
\begin{equation}
\label{eq:quantum-state}
\ket \psi = \sum_{x \in \F_2^n} a_x \ket x
\end{equation}
where $a_x \in \C$, for all $x \in \F_2^n$, and 
$\sum_{x\in \F_2^n}\abs{a_x}^2 = 1$. 
Suppose the the input state of $\mathfrak C'$ is an element of the standard basis, say $\ket x$ where $x = (x_1, \ldots, x_n) \in \F_2^n$. Below we describe the action of each individual gate \cite{NC} from the IBM-Q gate set described in Section~\ref{ibm}. \\
\begin{enumerate}
\item Action of $Z = \begin{bmatrix} 1 & 0 \\ 0 & -1 \end{bmatrix}$ on the 
$i^{th}$ qubit of $\ket x$: If there is a single $Z$ gate acting on the $i^{th}$ qubit, then the resulting transformation on a standard basis element is:
\begin{eqnarray}
\label{eq:actZ}
\begin{split} 
Z_i \ket{x_1 \ldots x_n} & = I\otimes \ldots \otimes Z \otimes \ldots \otimes I \ket{x_1 \ldots x_i \ldots x_n}\\
\mbox{i.e., } Z_i \ket{x} & = (-1)^{x_i} \ket{x_1 \ldots x_i \ldots x_n}\\
Z_i \ket{x} & =  I\otimes \ldots \otimes Z \otimes \ldots \otimes I \ket{x} = (-1)^{x_i} \ket{x}.
\end{split}
\end{eqnarray}

Thus the polynomial corresponding to the $Z$ gate is $f_{\mathfrak{C}}(x) = \exp{(\frac{2\pi\imath}{8} 4x_i)}$.\\

\item Action of 
$S = \begin{bmatrix} 1 & 0 \\ 0 & \exp(\frac{\imath \pi}{2}) \end{bmatrix}$ 
on the $i^{th}$ qubit of $\ket x$: If there is a single $S$ gate acting on the $i^{th}$ qubit, then the resulting transformation on a standard basis element is 

\begin{eqnarray}
\label{eq:actS}
\begin{split} 
S_i \ket{x_1 \ldots x_n} & = I\otimes \ldots \otimes S \otimes \ldots \otimes I \ket{x_1 \ldots x_i \ldots x_n} \\
&= \exp(\frac{\imath \pi x_i}{2}) \ket{x_1 \ldots x_i \ldots x_n}\\
\mbox{i.e., }S_i \ket{x} & = I\otimes \ldots \otimes S \otimes \ldots \otimes I \ket{x} \\ 
& = \exp(\frac{\imath \pi x_i}{2})\ket{x}.
\end{split}
\end{eqnarray}

Thus the polynomial corresponding to the $S$ gate is $f_\mathfrak{C}(x) = \exp{(\frac{2\pi\imath}{8} 2x_i)}.$ \\

\item Action of 
$T = \begin{bmatrix} 1 & 0 \\ 0 & \exp(\frac{\imath \pi}{4}) \end{bmatrix}$ 
on the $i^{th}$ qubit of $\ket x$: If there is a single $T$ gate acting on the $i^{th}$ qubit, then the resulting transformation on a standard basis element is 

\begin{eqnarray}
\label{eq:actT}
\begin{split} 
T_i \ket{x_1 \ldots x_n}& =I\otimes \ldots \otimes T \otimes \ldots \otimes I \ket{x_1 \ldots x_i \ldots x_n} \\
& = \exp(\frac{\imath \pi x_i}{4}) \ket{x_1 \ldots x_i \ldots x_n}\\
\mbox{i.e., }T_i \ket{x} & =I\otimes \ldots \otimes T \otimes \ldots \otimes I \ket{x} \\
& = \exp(\frac{\imath \pi x_i}{4})\ket{x}.
\end{split}
\end{eqnarray}

Thus the polynomial corresponding to the $T$ gate is $f_\mathfrak{C}(x) = \exp{(\frac{2\pi\imath}{8} x_i)}.$ \\

\item Action of 
$CZ = \begin{bmatrix} 1 & 0 & 0 & 0 \\ 
                               0 & 1 & 0 & 0 \\
                               0 & 0 & 1 & 0 \\
                               0 & 0 & 0 & -1  
\end{bmatrix}$ 
on the $i^{th}$ and $j^{th}$ qubits of $\ket x$:
If there is a single $CZ$ gate acting on the $i^{th}$ and $j^{th}$ qubits, then the resulting transformation on a standard basis element is 

\begin{eqnarray}
\label{eq:actCZ}
\begin{split} 
CZ_{ij}\ket{x_1 \ldots x_n} & = (-1)^{x_ix_j} \ket{x_1 \ldots x_i \ldots x_n}\\
\mbox{i.e., } CZ_{ij}\ket{x} & =  (-1)^{x_ix_j} \ket{x}.
\end{split}
\end{eqnarray}

$CZ_{ij}$ is a unitary transformation over 
$(\C^2)^{\otimes n}$ which acts as an identity operation on all qubits
of $\ket{x}$ except the $i^{th}$ and the $j^{th}$ qubit. It has the effect of 
a $CZ$ transform on the $i^{th}$ and the $j^{th}$ qubit.  
The above result is independent of the choice of target and the 
control qubits. 

Thus the polynomial corresponding to the $CZ$ gate is $f_\mathfrak{C}(x) = \exp{(\frac{2\pi\imath}{8} 4x_ix_j)}.$ \\

\item Action of 
$H = \frac{1}{\sqrt{2}} \begin{bmatrix} 1 & 1 \\ 1 & -1 \end{bmatrix}$ 
on the $i^{th}$ qubit of $\ket x$: The unitary transformation on $\ket x$ 
which keeps all the qubits except the $i^{th}$ unchanged and applies 
$H$ on the $i^{th}$ qubit is denoted by $H_i$. 
Bremner, Jozsa and Shepher \cite{Bremner} have proved that 
by introducing an ancilla qubit $\ket{x_a}$, and thus expanding the 
input state $\ket x$ to $\ket x  \ket{x_a}$, 
$H_i$ can be replaced by the operator $H_i CZ_{ai}H_a$. 
Then, the following is obtained.
\begin{equation}
\label{eq:H}
H_i CZ_{ai}H_a \ket x \ket{x_a}= \frac{(-1)^{x_i x_a}}{\sqrt{2}} \ket x \ket{x_a}. 
\end{equation}
\begin{figure}[h]
\begin{center}
\includegraphics[scale=0.55]{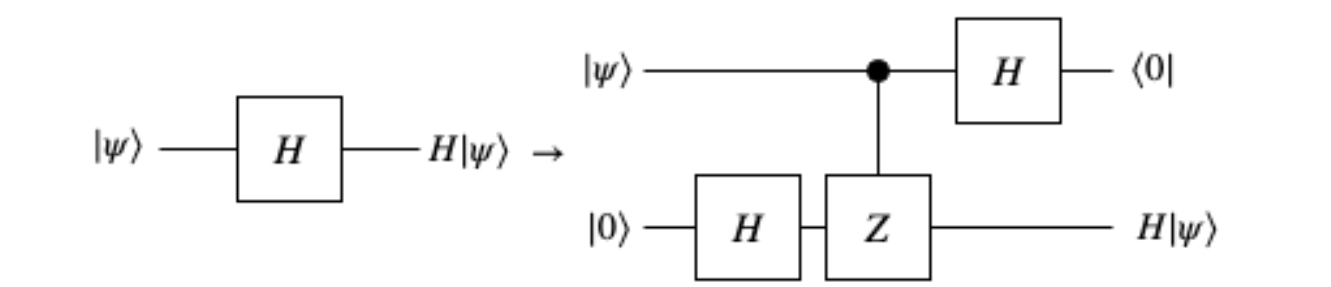}
\caption{Hadamard Gadget}
\label{fig:Hadamard_G}
\end{center}
\end{figure}
Let $\ket{x_i}$ be state of the $i^{th}$ qubit on which a Hadamard transform 
is applied. We introduce an ancilla qubit $\ket{x_a}$ where $x_i, x_a \in \F_2$.  
Since the other qubits remain unchanged by the transformation,
without any loss of generality we consider the action of $H_i CZ_{ai}H_a$ on 
$\ket{x_i} \ket{x_a}$. It is to be noted that if we consider the quantum 
state $\ket{x_i}\ket{x_a}$, then $H_a = I \otimes H$ and $H_i = H \otimes I$.

\begin{eqnarray*}
\begin{split}
\ket{x_i} \ket{x_a} & \xmapsto{H_a}
\frac{1}{\sqrt{2}}(\ket{x_i}\ket{0} + (-1)^{x_a}\ket{x_i}\ket{1})  \\
& \xmapsto{CZ_{ia}} \frac{1}{\sqrt{2}}( \ket{x_i}\ket{0} + (-1)^{x_i + x_a}\ket{x_i}\ket{1})  \\
&\xmapsto{H_i} \frac{1}{2}((\ket{0} + (-1)^{x_i}\ket{1})\ket{0} \\
& \qquad \qquad + (-1)^{x_i + x_a}(\ket{0} + (-1)^{x_i}\ket{1})\ket{1})\\
&= \frac{1}{2} \ket{0}\left( \ket{0} + (-1)^{x_i+x_a}\ket{1}\right) \\
& \qquad \qquad + \frac{1}{2}\ket{1}\left( (-1)^{x_i} \ket{0} + (-1)^{x_a}\ket{1}\right)\\
&= \frac{1}{\sqrt{2}}\ket{0}
\left( \frac{\ket{0} + (-1)^{x_i+x_a}\ket{1}}{\sqrt{2}} \right) \\
& \qquad + \frac{1}{\sqrt{2}}(-1)^{x_i}\ket{1}
\left( \frac{\ket{0} + (-1)^{x_i +x_a}\ket{1}}{\sqrt{2}} \right).
\end{split}
\end{eqnarray*}

Summing up, the net effect of $H_iCZ_{ia}H_a$ on $\ket{x_i}\ket{x_a}$ is 

\begin{eqnarray*}
\begin{split}
\ket{x_i} \ket{x_a}  &\xmapsto{H_iCZ_{ia}H_a} 
\frac{1}{\sqrt{2}}\ket{0}
\left( \frac{\ket{0} + (-1)^{x_i+x_a}\ket{1}}{\sqrt{2}} \right)\\
& \qquad +
\frac{1}{\sqrt{2}}(-1)^{x_i}\ket{1}
\left( \frac{\ket{0} + (-1)^{x_i +x_a}\ket{1}}{\sqrt{2}} \right). 
\end{split}
\end{eqnarray*}
Setting $x_a = 0$
\begin{eqnarray*}
\begin{split}
\ket{x_i} \ket{0}  &\xmapsto{H_iCZ_{ia}H_a} 
\frac{1}{\sqrt{2}}\ket{0}
\left( \frac{\ket{0} + (-1)^{x_i}\ket{1}}{\sqrt{2}} \right)\\
& \qquad\qquad +
\frac{1}{\sqrt{2}}(-1)^{x_i}\ket{1}
\left( \frac{\ket{0} + (-1)^{x_i}\ket{1}}{\sqrt{2}} \right). 
\end{split}
\end{eqnarray*}

Thus, we observe that if $x_a = 0$ and post-selection is performed on $\ket{0}$, then the ancilla qubit will 
contain the Hadamard transform of $\ket{x_i}$. If we decide to 
neglect the $i^{th}$ qubit and consider only the ancilla qubit after the 
transformation then we will have the same effect of a Hadamard
gate acting on the $i^{th}$ qubit. 
gate.
Since we know that we can neglect the first and the last Hadamard gates applied to any qubit, we neglect the first on the ancilla qubit and the corresponding last on the $i^{th}$ qubit as it is considered one pathway in the quantum circuit. Therefore, the only transformation we have to consider is the $CZ_{ia}$ 
gate.
Thus, 
\begin{equation*}
\ket{x}\ket{x_a} \mapsto \frac{(-1)^{x_i x_a}}{\sqrt{2}} \ket{x}\ket{x_a}.  
\end{equation*}

Thus the polynomial corresponding to the $H$ gate is $f_\mathfrak{C}(x) = \exp{(\frac{2\pi\imath}{8} 4x_ix_j)}$. The normalizing factor of $\frac{1}{\sqrt{2}}$ is incorporated in the polynomial for the entire circuit. \\

\item Action of $X = \begin{bmatrix} 0 & 1 \\ 1 & 0 \end{bmatrix}$ on the 
$i^{th}$ qubit of $\ket x$: If there is a single $X$ gate acting on the $i^{th}$ qubit, then the resulting transformation can be represented by the transformation carried by the gates $HZH$.
\begin{figure}[h]
\begin{center}
\includegraphics[scale=0.65]{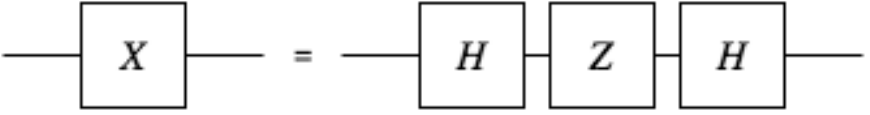}
\caption{X gate decomposed in terms of Z and H gates, where $X = HZH$}
\label{fig:X}
\end{center}
\end{figure}

\begin{eqnarray*}
\label{eq:actX}
\begin{split} 
X_i \ket{x_1 \ldots x_n} & = I\otimes \ldots \otimes X \otimes \ldots \otimes I \ket{x_1 \ldots x_i \ldots x_n} \\
& = I\otimes \ldots \otimes HZH \otimes \ldots \otimes I \ket{x} \\
\mbox{i.e., } X_i \ket{x} & =  I\otimes \ldots \otimes H_iCZ_{a_1i}H_{a_1}ZH_iCZ_{a_2i}H_{a_2} \\
& \qquad \quad \quad \quad \otimes \ldots \otimes I \ket{x}\ket{a_1}\ket{a_2}\\
\mbox{i.e., } X_i \ket{x} & = \frac{1}{2}\e^{\iota\frac{\pi}{4}{4(x_ia_1 + a_1 + a_1a_2)}}\ket{x}.
\end{split}
\end{eqnarray*}

Thus the polynomial corresponding to the $X$ gate is $f_\mathfrak{C}(x) = \frac{1}{2}\exp{\frac{2\pi\imath}{8} 4(x_ia_1 + a_1 + a_1a_2)}.$ \\

\item Action of $CNOT = \begin{bmatrix}      1&0&0&0\\ 0&1&0&0\\ 0&0&0&1\\ 0&0&1&0\\ \end{bmatrix}$ on the 
$i^{th}$ and $j^{th}$ qubit of $\ket x$ (where $i^{th}$ qubit is the control qubit and the $j^{th}$ qubit is the target qubit: If there is a single $CNOT$ gate acting on the $i^{th}$ and $j^{th}$ qubit, then the resulting transformation can be represented by the transformation carried by the gates $H_jCZ_{ij}H_j$.

\begin{figure}[H]
\begin{center}
\includegraphics[scale=0.55]{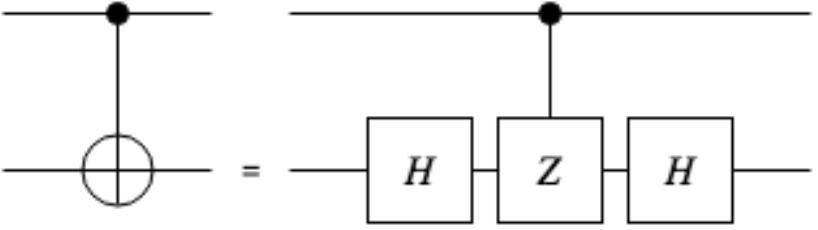}
\caption{CNOT gate decomposed in terms of H and CZ gates, where $CNOT = H_jCZ_{ij}H_j$}
\label{fig:CNOT}
\end{center}
\end{figure}

\begin{eqnarray*}
\label{eq:actCNOT}
\begin{split} 
CNOT_{ij} \ket{x_1 \ldots x_n} &= I\otimes \ldots \otimes CNOT \otimes \ldots \\ 
& \qquad \quad \ldots \otimes I \ket{x_1 \ldots x_i \ldots x_n} \\
CNOT_{ij} \ket{x} &=  I\otimes \ldots \otimes -\iota H_jCZ_{ij}H_j \otimes \ldots \\ & \qquad \quad \ldots \otimes I \ket{x}\\
CNOT_{ij} \ket{x} &=  I\otimes \ldots \otimes -\iota H_jCZ_{a_1j}H_{a_1}CZ_{ij} \\ 
& H_jCZ_{a_2j}H_{a_2} \otimes \ldots \otimes I \ket{x}\ket{a_1}\ket{a_2}\\
CNOT_{ij} \ket{x} &= \frac{1}{2}\e^{\iota\frac{\pi}{4}{4(x_ja_1 + x_ia_1 + a_1a_2)}}\ket{x}.
\end{split}
\end{eqnarray*}

Thus the polynomial corresponding to the $CNOT$ gate is $f_\mathfrak{C}(x) = \exp{\frac{2\pi\imath}{8} 4(x_ja_1 + x_ia_1 + a_1a_2)}.$ \\

\item Action of $Y = \begin{bmatrix} 0 & -\iota \\ \iota & 0 \end{bmatrix}$ on the 
$i^{th}$ qubit of $\ket x$: If there is a single $Y$ gate acting on the $i^{th}$ qubit, then the resulting transformation can be represented by the transformation carried by the gates $-\iota ZHZH$.

\begin{figure}[H]
\begin{center}
\includegraphics[scale=0.6]{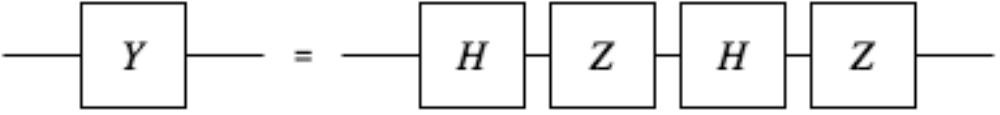}
\caption{Y gate decomposed in terms of Z and H gates, where $Y = -\iota ZHZH$}
\label{fig:Y}
\end{center}
\end{figure}

\begin{eqnarray*}
\label{eq:actY}
\begin{split} 
Y_i \ket{x_1 \ldots x_n} &= I\otimes \ldots \otimes Y \otimes \ldots \otimes I \ket{x_1 \ldots x_i \ldots x_n}\\
\mbox{i.e., } Y_i \ket{x} &=  I\otimes \ldots \otimes -\iota Z_iH_iZ_iH_i \otimes \ldots \otimes I \ket{x}\\
\mbox{i.e., } Y_i \ket{x} &=  I\otimes \ldots \otimes -\iota Z_iH_iCZ_{a_1i}H_{a_1}Z_iH_i\\
& \qquad \quad CZ_{a_2i}H_{a_2} \otimes \ldots \otimes I \ket{x}\ket{a_1}\ket{a_2}\\
\mbox{i.e., } Y_i \ket{x} &= \frac{-1}{2}\iota\e^{\iota\frac{\pi}{4}{4(x_i a_1 + a_1 + a_1a_2 + a_2)}}\ket{x}.
\end{split}
\end{eqnarray*}

Thus the polynomial corresponding to the $Y$ gate is $f_\mathfrak{C}(x) = \frac{-1}{2}\exp{\frac{2\pi\imath}{8} 4(x_ia_1 + a_1 + a_1a_2 + a_2+ \frac{1}{2})}.$ \\

\end{enumerate}


\section{Writing polynomial corresponding to a quantum circuit}
Based on the analysis done in the last section, we can write a general procedure to find the polynomial corresponding to a quantum circuit.
\subsection{Rules}
Below we describe the rules that are followed while forming the polynomial associated with a quantum circuit. These rules are based on the results shown above.

\begin{itemize}
    \item Assign input variables to the leftmost point of all the wires sequentially.
    \item Traverse the top wire from left to right and apply the following transformations based on the quantum gate encountered:
    \begin{itemize}
        \item $Z$ gate: Add 4 $\times$ (input variable) to the polynomial.
        \item $S$ gate: Add 2$\times$(input variable) to the polynomial.
        \item $T$ gate: Add 1$\times$(input variable) to the polynomial.
        \item $H$ gate: Introduce a new ancilla variable as the output and add the 2-degree term 4$\times$(input variable)$\times$(output ancilla variable) to the polynomial.
        \item $CZ$ gate: Add 4$\times$(control input variable)$\times$(target input variable) to the polynomial.
        \item $CCZ$ gate: Add 4$\times$(control input variable 1)$\times$(control input variable 2)$\times$(target input variable) to the polynomial.
        \item For any other gate, the resultant term to be added to the polynomial is simply a combination of the gates above. The rules for $X$, $Y$ and $CNOT$ have been \textbf{derived} using the above rules.
    \end{itemize}
\end{itemize}
\subsection{Example}
\begin{figure}[H]
\begin{center}
\includegraphics[scale=0.55]{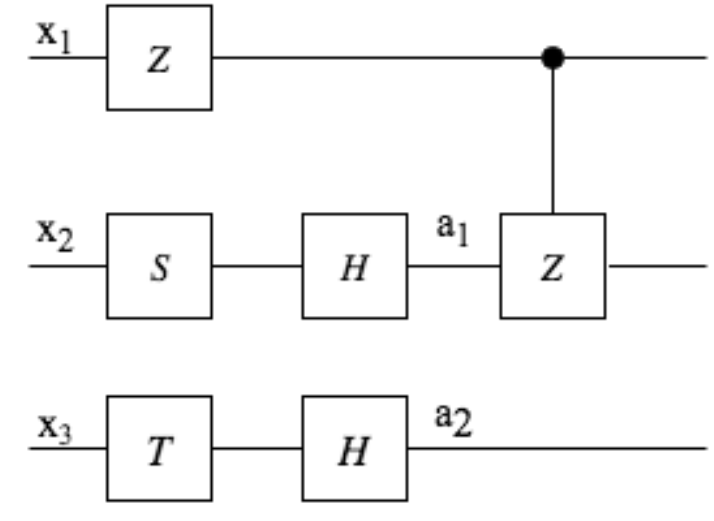}
\caption{The polynomial $f_\frak{C}$ of this sample circuit is $4x_1 + 2x_2 + 4x_2a_1 + 4x_1a_1 + x_3 + 4x_3a_2$}
\label{fig:Sample Circuit}
\end{center}
\end{figure}

\noindent The rules can be understood by observing the polynomial corresponding to this example circuit. In figure \ref{fig:Sample Circuit},
\begin{itemize}
    \item At the wire's leftmost point, the topmost qubit is assigned the variable $x_1$, middle qubit $x_2$ and bottom-most qubit $x_3$.
    \item Traverse the top wire from left to right
    \begin{itemize}
        \item For the Z gate add the term $4x_1$ to the polynomial.
        \item For the $CZ$ gate as we are not aware of the variable associated with the target qubit, we leave its processing for later.
    \end{itemize}
    \item Traverse the wire associated with the variable $x_2$
    \begin{itemize}
        \item For the $S$ gate add the term $2x_2$ to the polynomial.
        \item For the $H$ gate add $4x_2a_1$, where $a_1$ is the ancilla qubit introduced by this $H$ gate.
        \item Observe the input to the target qubit of the aforementioned $CZ$ gate to be $a_1$. Hence add the term $4x_1a_1$ to the polynomial.
    \end{itemize}
    \item Traverse the wire associated with the variable $x_3$.
    \begin{itemize}
        \item For the $T$ gate add the term $x_3$ to the polynomial.
        \item For the $H$ gate add the term $4x_3a_2$ to the polynomial where $a_2$ is the ancilla qubit introduced by the $H$ gate.
    \end{itemize}
\end{itemize}
         Thus, the net polynomial associated with the circuit is thus,
\begin{equation*}
    f_\frak{C}(x) = 4x_1 + 2x_2 + 4x_2a_1 + 4x_1a_1 + x_3 + 4x_3a_2
\end{equation*}

\section{Generalized Walsh-Hadamard transform and quantum circuits}

In the last section we have seen that given any quantum circuit 
$\frak C$ involving $n$ qubits, it is possible to covert it to a circuit
involving $n + m$ qubits, by introducing $m$ ancilla qubits, of the 
form $H^{\otimes (n+m)}\frak C' H^{\otimes (n+m)}$, where the 
inner circuit $\frak C'$ contains only diagonal gates, and its action 
on a standard basis vector $\ket x$ is given by 


\begin{equation*}
\frak C' \ket x \xmapsto{\frak C'} 
\exp\left(\frac{2\pi \imath f_{\frak C'}(x)}{8}\right)\ket x
\end{equation*}
where $x \in \mathbb \{0, 1\}^{n+m}$. The discussion in the previous
section tells us that $f_{\frak C'}$ is of the form
\begin{equation*}
f_{\frak C'}(x) = a_0 + \sum_{i \in [n+m]} a_i x_i + \sum_{i, j \in [n+m], i < j} a_{ij}x_i x_j
\end{equation*}
where $a_0, a_i, a_{ij} \in \Z_8$, for all $i, j \in [n+m]$. Let $0_n \in \F_2^n$ 
be the vector with all coordinates equal to $0$.

\begin{eqnarray*}
\label{circuit}
\begin{split}
\ket{0_{n+m}} &\xmapsto{H^{\otimes (n+m)}} 
2^{-\frac{n+m}{2}}\sum_{x\in \F_2^{n+m}} \ket x \\
& \xmapsto{\frak C'}  2^{-\frac{n+m}{2}}\sum_{x\in \F_2^{n+m}} \exp\left(\frac{2\pi \imath f_{\frak C'}(x)}{8}\right)\ket x \\
&\xmapsto{H^{\otimes (n+m)}}
2^{-(n+m)}\sum_{x\in \F_2^{n+m}} \exp\left(\frac{2\pi \imath f_{\frak C'}(x)}{8}\right). \\ 
& \qquad \qquad \qquad \qquad \quad \sum_{y\in \F_2^{n+m}}(-1)^{x \cdot y} \ket y \\
&= 2^{-(n+m)}\sum_{y\in \F_2^{n+m}} 
\sum_{x\in \F_2^{n+m}}\exp\left(\frac{2\pi \imath f_{\frak C'}(x)}{8}\right).\\ 
& \qquad \qquad \qquad \qquad \qquad \qquad (-1)^{x \cdot y} \ket y\\ \\
&= 2^{-(n+m)}\sum_{y\in \F_2^{n+m}} 
\sum_{x\in \F_2^{n+m}}\zeta_8^{f_{\frak C'}(x)} (-1)^{x \cdot y} \ket y \\
& = 2^{-(n+m)}\sum_{y\in \F_2^{n+m}} \mathcal H_{f_{\frak C'}}^{(8)}(y)\ket y\\
\end{split}
\end{eqnarray*}

The relation between the probability amplitudes of the output quantum 
state of the circuit $H^{\otimes (n+m)}\frak C' H^{\otimes (n+m)}$ 
to the generalized Walsh-Hadadamard spectrum of 
$f_{\frak C'}: \F_2^{n+m} \rightarrow \Z_8$ 
is explicitly derived in \eqref{circuit}. It is also evident that $f_{\frak C'}$ can 
be associated to a quadratic form over $\Z_8$. Analyzing these spectra obtained
from IBM-Q quantum algorithms provides a strong motivation to study 
generalized Boolean functions. 

\section{Implementation of Swap and Toffoli gate on IBM-Q and the associated polynomial}
The swap gate can be represented in the polynomial form as in the below figure.  

\begin{figure}[H]
\begin{center}
\includegraphics[scale=0.35]{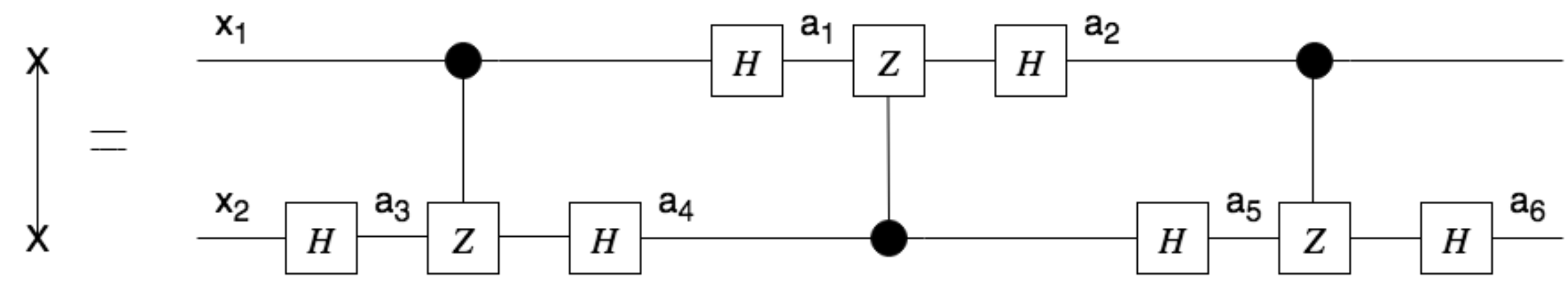}
\caption{The polynomial for the swap gate is $f_\frak{C}(x) = 4(x_1a_3 + x_1a_1 + a_1a_2 + a_2a_5 + x_2a_3 + a_3a_4 + a_4a_1 + a_4a_5 + a_5a_6$)}
\label{fig:swap_gate}
\end{center}
\end{figure}

\noindent
The Toffoli gate is represented in polynomial form as in the below figure.

\begin{figure}[H]
\begin{center}
\includegraphics[scale=0.32]{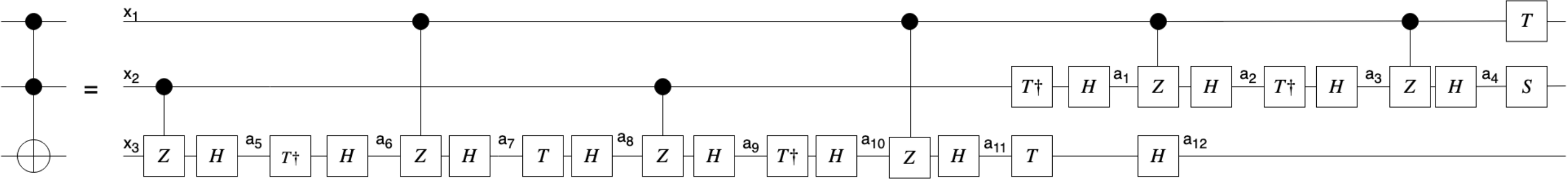}
\caption{The polynomial for the Toffoli gate is $f_\frak{C}(x) = x_1 - x_2 - a_2 - a_5 + a_7 - a_9 + a_{11} + 2a_4 + 4(x_1a_6 + x_1a_{10} + x_1a_1 + x_1a_3 + x_2x_3 + x_2a_8 + x_2a_1 + a_1a_2 + a_2a_3 + a_3a_4 + x_3a_5 + a_5a_6 + a_6a_7 + a_7a_8 + a_8a_9 + a_9a_{10} + a_{10}a_{11} + a_{11}a_{12})$}
\label{fig:toffoli_gate}
\end{center}
\end{figure}

\section{Conclusion}
In this work, we expand polynomial representations of quantum circuits to the complex domain by providing a polynomial representation of the Y, S, T gates. This allows us to represent a large number of quantum circuits as generalized Boolean polynomials. One of the limitations of this approach is that a polynomial can correspond to more than one quantum circuits. Nevertheless, by studying the generated polynomials, we feel that it will provide insights into quantum circuit design and their reduction. Also, by studying alternative representations of quantum gates, perhaps a general and more efficient procedure for converting circuits to polynomials can be developed.
In future, the aim is to compute the Boolean polynomial corresponding to a quantum circuit that suffers from gate and readout errors due to various noise sources. Subsequently, by using polynomial algebra, we hope to develop a method to minimize error in performing a given computational task on such a noisy circuit.

\section{Acknowledgment}
This research is funded by QuLabs Software
India Pvt. Ltd. through the project
QuLabs@IITR. 

\end{document}